\documentclass[lettersize,journal]{IEEEtran}
\usepackage{amsmath,amsfonts}
\usepackage{algorithm}
\usepackage{algpseudocode}
\usepackage{amssymb}

\usepackage{mathtools}
\usepackage{float}
\usepackage{url}
\usepackage{linegoal}
\usepackage{amsmath}
\usepackage{varioref}
\usepackage{tipa}
\usepackage{times}
\usepackage{xspace}
\usepackage{graphicx}
\usepackage{tabularx}
\usepackage{booktabs}
\usepackage{multirow}
\usepackage[margin=1cm]{subcaption}
\usepackage{threeparttable}

\title{Co-Simulation Framework For Network Attack Generation and Monitoring}

\author{\IEEEauthorblockN{Oceane Bel, Joonseok Kim, William J Hofer, Manisha Maharjan, Sumit Purohit, Shwetha Niddodi}\\
\thanks{The research described in this paper is part of the Resilience Through Data-Driven, Intelligently Designed Control (RD2C) Initiative at Pacific Northwest National Laboratory. It was conducted under the Laboratory Directed Research and Development Program at PNNL, a multiprogram national laboratory operated by Battelle for the U.S. Department of Energy. 

O. Bel, W. Hofer, M. Maharjan, S. Purohit and S. Niddodi are with Pacific Northwest National Laboratory, Richland, WA, 99352, USA (e-mail: \{obel@pnnl.gov, william.hofer, manisha.maharjan, sumit.purohit, shwetha.niddodi\}@pnnl.gov).

J.-S. Kim is with the National Security Sciences Directorate, Oak Ridge National Laboratory, Oak Ridge, TN 37830, USA (e-mail: kimj1@ornl.gov). 
}
}

\date{September 2022}

\newcommand{\dnp}{DNP3\xspace}
\newcommand{\ns}{NS3\xspace}
\newcommand{\gridlabd}{GridLAB-D\xspace}
\newcommand{\helics}{HELICS\xspace}
\newcommand{\MIM}{man-in-the-middle\xspace}
\newcommand{\docker}{Docker\xspace}
\newcommand{\wifi}{Wi-Fi\xspace}
\begin{document}

\maketitle
\begin{abstract}
    Resilience assessment is a critical requirement of a power grid to maintain high availability, security, and quality of service. Most grid research work that is currently pursued does not have the capability to have hardware testbeds. Additionally, with the integration of distributed energy resources, the attack surface of the grid is increasing. This increases the need for reliable and realistic modeling techniques that are usable by the wider research community. Therefore, simulation testbeds have been used to model a real-world power grid topology and measure the impact of various perturbations.
    
    Existing co-simulation platforms for powergrid focus on a limited components of the overall system, such as focusing only on the dynamics of the physical layer. Additionally a significant number of existing platforms need specialized hardware that may be too expensive for most researchers. Finally, not many platforms support realistic modeling of the communication layer, which requires use of Supervisory Control and Data Acquisition communication protocol such as \dnp while modeling cybersecurity scenarios. 
    
    We present Network Attack Testbed in [Power] Grid (NATI[P]G), (pronounced \textit{natig}), a standalone, containerized, and reusable environment to enable cyber analysts and researchers to run different cybersecurity and performance scenarios on powergrid. Our tool combines GridLAB-D, a grid simulator, \helics, a co-simulation framework, and NS-3, a network simulator, to create an end-to-end simulation environment for the power grid. We demonstrate use cases by generating a library of datasets for several scenarios. These datasets can be used to detect cyberattacks at the cyber layer, and develop counter measures to these adverse scenarios.  
    
\end{abstract}

\section{Introduction}
Cyber-physical systems (CPS), such as microgrids, are key infrastructure components that impact social, financial, and national security on a daily basis. Thus, with cyberattacks increasing in sophistication by the day, there is a need to understand various failure scenarios and plan for mitigation strategies for a reliable and resilient power grid operation \cite{lee2015electric, dutta2022cyber}. Simulation environments have been extensively used to model CPS components, topologies, adversaries, attack sequences, and their impact on the systems~\cite{zografopoulos2021cyber,bhattacharya2020automated}. CPSs are complex and interdependent systems, with both discrete and continuous measurements. However, existing attack detection models, such as data-driven  intrusion  detection  and  prevention  systems, require high amount of data to train models before they are deployed on the grid. This stymies efforts to combat the rate of improvement that cyberattackers have when attacking Cyber-physical systems.

To develop adequate defenses, we must efficiently generate end-to-end models of systems and adversaries. Current simulators, such as NetSim~\cite{barnett1992netsim} and Mininet~\cite{lantz2010network}, provide a partial view of the system, but fail to exhibit physical constraints and conditional operations across different components. Co-simulation environments have been developed to address these limitations, but struggle to address usability and flexibility challenges. Additionally, the co-simulators provide little or no support for \textit{perturb-and-observe} to model adversarial scenarios and generate benchmark datasets for downstream applications such as risk assessment, attack detection, and risk mitigation.

We present Network Attack Testbed in [Power] Grid (NATI[P]G), a co-simulation environment for distribution power grid network using state-of-the-art simulators. This co-simulator is used to generate attack scenarios that can enable researchers to understand how attackers could behave in a network given a set of goals. Our work builds upon past work where researchers modeled attacks using network simulators to understand the behavior of attackers~\cite{lee2014simulated,ashok2021high}. By modeling potential adversary behaviors, researchers can develop faster ways to identify them on the network during attacks. 

We focus on \MIM attacks on a power grid as our primary attack scenario. We implemented these attacks on the \ns network simulator and measure their impact in the \gridlabd simulator. We do not limit our scenarios to transport and session layers, but rather demonstrate application layer perturbations in the communication layer. The goal is to provide an example on how our tool can be used by other researchers without specialized hardware. Using our testbed, we identified different behaviors between grid following and grid forming inverters during cyberattacks. We also use our testbed to find settings for capacitors and generators to minimize frequency deviation when switches are tripped and microgrids are islanded.

The contributions of the paper are as follows:
\begin{itemize}
  \item Our co-simulation tackles the entire stack, creating a simulation close to what is expected of a real test bed.
  \item We produce a containerized framework for a wide range of cyber resilience assessment applications, improving upon existing testbeds.
  \item We demonstrate the feasibility of modeling, simulating, and validating CPS environments without using specialized hardware.
 \item We leverage application layer commands using \dnp protocol in \ns to simulate realistic grid behaviour.
  \item We simulate \MIM scenarios using \dnp protocols.
\end{itemize}

This paper is organized as follows. In Section 2, we survey existing approaches and address our motivation. Section 3 delineates the design of our framework and use cases of cyberattacks. Section 4 describes the details of attack scenarios to demonstrate the usability of our framework. We report our experimental results in Section 5. Section 6 concludes our work and discusses future work.

\section{Background and related work}
The power grid is a critical infrastructure due to its effect on daily life~\cite{dabrowski2017grid}. Energy providers must balance supply with demand and handle unforeseen events, such as extreme weather patterns. These events can affect the functionality of the grid and impact a large portion of the population. Any threats to the grid should be identified early enough so that providers can deal with the threats before they impact the functions of the grid~\cite{aoufi2020survey}. 

In recent years, traditional power systems have become more integrated with information and communication technology, which has given way to Cyber-Physical Power Systems~\cite{yohanandhan2020cyber}. Alongside the growth of network simulators, more researchers are turning to simulators to develop new technology. This means that simulators must evolve to get closer to realistic networks without losing simulation efficiency.

\subsection{\ns}

Network Simulator 3 (\ns)~\cite{carneiro2010ns} is a network simulator commonly used in network architecture and system development. It can simulate almost all aspects of a network, including the physical proprieties of signal transmission. This allows users greater control over the different aspects of the network, allowing them to simulate various attack scenarios with different network topologies, interject data flows at a particular node, and replace normal data with false data. An example of this control is setting whether or not a packet has reached its final destination. This simulates how an attacker can stop a packet at the node controlled by them before sending out a new packet with different information to a victim node. Another example is updating a source IP to the original source IP that was used by the intercepted packet, leaving a victim unaware of an IP change.  

Using the simulator, we can create datasets and a system that can be used by other researchers to model potential attacks. Our system has a configuration file that can be used to set different topologies and tune attack parameters. The simulator can also collect performance and routing information using \ns monitors. The monitored information can be used to create attacker models that can be used as controllers in a network to adapt the network settings in response to attackers. 

\subsection{\gridlabd}
GridLAB-D is a simulation tool that enables power distribution system simulation and analysis~\cite{chassin2008gridlab}. It is extensively used to represent the system behavior of different power system components and complex interactions between these components and modern grid technologies like distributed energy resources (DERs). It can perform power flow analysis and dynamic studies and generates detailed load and market models. With the recent interests in studies regarding the interaction of power system and communication networks, \gridlabd has been used for bench-marking IEEE feeder models for modeling detailed and dynamic power system operations, and time-series power flow simulations~\cite{bhattarai2020studying,guddanti2020better}. It facilitates larger simulations of power system models and simplified implementation of system architectures for illustrating a wide range of scenarios that reflect the impacts of communication layers on power system operations.

\subsection{Co-simulation Environment}
There are existing platforms that simulate power systems. One of these platforms~\cite{mana2020study} includes a mechanism that uses \ns as a networking interface in conjunction with Framework for Network Co-Simulation (FNCS) and \gridlabd as tools to keep track of value changes over time of power system components. In line with this work, we utilize a co-simulation to simulate cyberattacks on power networks. Additionally, our tool allows for larger simulation by using \helics instead of FNCS as the interconnect between \gridlabd and \ns3. Battarai et al.~\cite{bhattarai2020studying} presents a \helics based co-simulation environment, but the work does not focus on scenarios involving application layer perturbations. We introduce \dnp protocol as part of \ns to send measurement data and control commands between the \ns nodes representing components in SCADA systems.

\subsection{Reconfigurable networks}
Current networks have the ability to reconfigure themselves as workloads and needs change. Several things can happen during reconfiguration, such as user equipment changing which antenna it connects to or topology changes. Slicing, in 5G networks, is another way to implement reconfiguration, where partitioning the network can be done to isolate network attacks. This prevents an attacker from harming the entire network and spreading its influence. Slices can be updated as needed over time to isolate network sections or to distribute resources in response to a metric such as performance or monetary spending. Therefore, live reconfiguration can be a powerful tool for security and for enhancing performance enhancement of networks. A current research area revolves around how to make use of 5G network slicing and reconfigurability for power grids. 

\subsection{\dnp protocol}
\dnp is a protocol used to send packets between nodes, commonly in a utility distribution network~\cite{siddavatam2015security}. According to surveys, more than 75\% of North American electric utilities use or have used \dnp as a communication protocol~\cite{east2009taxonomy}. Using this protocol in conjunction with the \ns simulator allowed us to develop an end-to-end power grid traffic simulator. Using this setup, we can simulate different situations, such as downed nodes or attacks on the network. We can also monitor the traffic flow between the nodes to model any changes in the traffic between when the attack is happening and when the traffic is normal.

\subsection{Cybersecurity simulations platforms}
Previous work has been done with the goal of generating cyber attacks, such as \MIM attacks on LAN wireless network~\cite{kumar2012detection}, WAN networks~\cite{liu2015analyzing}, hard connected~\cite{apriani2020implementation} and VANET networks~\cite{mejri2016new}. Additionally, researchers have looked at the effect of cyberattacks, like \MIM and denial of service, on power systems, and have found that denial of service attacks has significant impacts on the run time of devices on the network~\cite{devanarayana2019testing}. We focus on creating a lightweight simulation tool that can be used by other researchers. The tool can be used to parameterize the attack that they want to simulate and get traffic information characterizing the effect of the attack on the grid. Another example of attack generation platform is GridSTAGE~\cite{nandanoori2021nominal}, which can be used to simulate false data injection attacks. It provides a framework where the user can input the parameters of an attack, and measure its effects on network traffic.

\section{Co-simulation Design}

This section briefly describes the co-simulation platform developed for simulating the cyber-physical dynamics of the distribution grid and simulating attack scenarios at various parts in the grid. The platform combines various industry-grade and open-source tools such as \gridlabd, \ns, \helics to emulate the different layers of the grid infrastructure. The platform uses the communication protocol, \dnp, for the grid communications. This paper focuses on generation of man-in-the-middle attack behavior in distribution grid. We use a \docker container to distribute our tool to other researchers. \docker was chosen since it allows us to create a lightweight environment that can easily be used by other researchers. The container enables researchers that don't have access to an real power network to conduct research on a realistic environment.

\begin{figure*}[!htb] 
  \centering
  \includegraphics[width=1\linewidth]{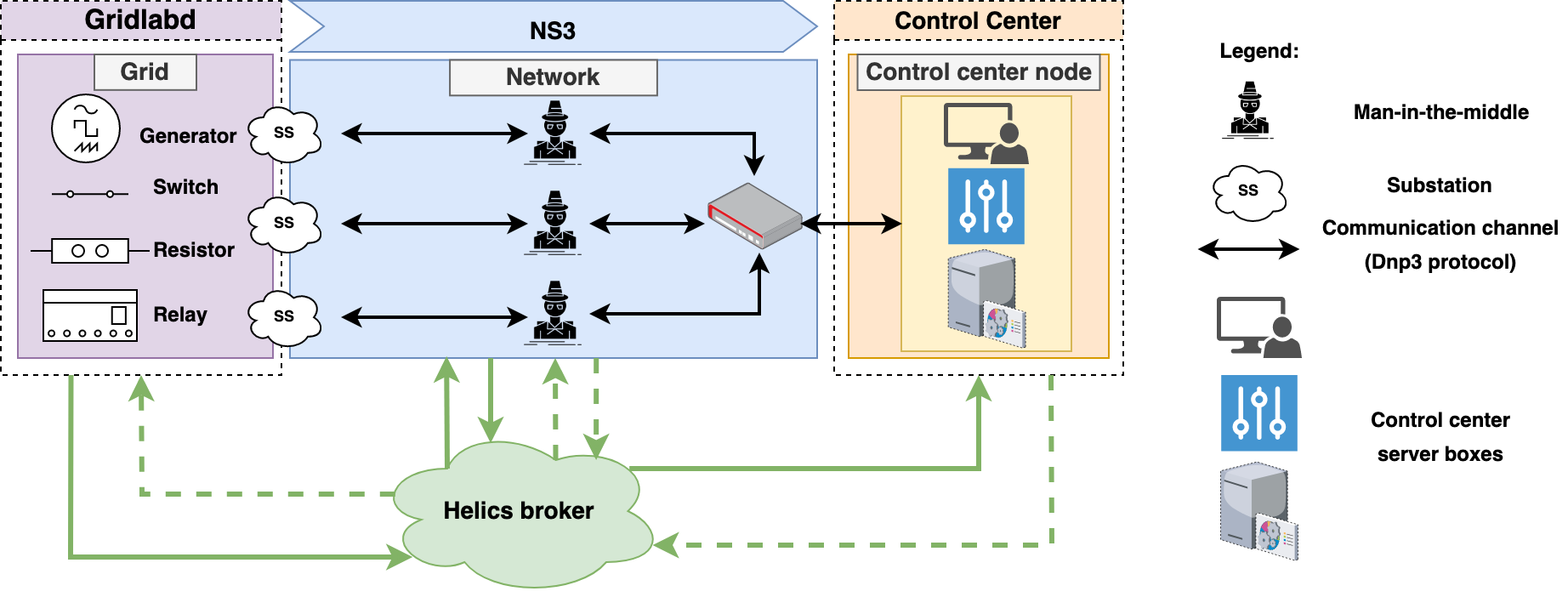}
  \caption{Overview of the co-simulation environment with interactions between \helics, \gridlabd and \ns. Each node uses the \dnp protocol to communicate. The control center, where the Open Platform Communications (OPC) server is located, is responsible for control a region of the grid network. The substation is responsible for power distribution and aggregating the information for the microgrid to be sent to the control center. The control center does periodic poll requests every 4 seconds. Once the poll request is received, the microgrid/substation responds with collected measurements. }
  \label{fig:environment}
\end{figure*}

\subsection{Our co-simulation platform}

Our current platform is divided into three layers: the control layer, the network or communication layer, and the physical layer. \gridlabd is used for the physical layer, while \ns is used to construct the connecting network. Finally the control layer is represented as the control node and it is controlled through the main simulation program. The control layer consists of one utility control center that receive measurements from and send control commands to the respective microgrids in the physical layer. The network layer represents the communication medium that carries information between the control and the physical layer. This layer consists of various network topology and uses \dnp as grid communication protocol between the control layer and physical layer. 

The physical layer consists of the physical distribution grid using IEEE 123 bus test feeder model with various DERs. The IEEE 123 bus feeder model is further logically split into three microgrids which can be configured in grid connected or islanded mode. Each microgrid has a substation which have remote terminal units (RTUs) that aggregate data from and disseminate control signals coming from the control center to various DERs in the microgrid. The control center does periodic poll requests every 4 seconds. Once the poll request is received, the microgrid/substation responds with collected measurements. A \MIM attack node sits between the control center and each of the substations. The \MIM attacker changes the data that are sent between the substations and the control centers. The attacker's goal is to trick users into thinking the substations sent good data to the control center while sending a different command to the substation. The different command can be used to collect additional information from the substation or modify parameters such as tripping a relay in the microgrid.

\subsubsection{\gridlabd and IEEE 123 node test feeder}
The GridLAB-D simulation tool is used to model the IEEE 123 node test feeder~\cite{guddanti2020better}. This test feeder is used as the base power system model for the developed tool and multiple distributed energy resources (DERs) are integrated at different locations to construct a distribution system architecture with three different microgrids. The microgrids can be connected to the grid or islanded in different combinations to create different feeder structures. Each microgrid consists of three DERs: one grid-following inverter based photovoltaics (PV), one grid-connected inverter based PV and one diesel generator. The generator is modeled using synchronous machine with simple excitation system enabling droop curve to the voltage/reactive power output, and GGOV1 governor model with primary power and frequency droop controls. 
Similarly, inverters are equipped with current and voltage control loops to adjust various droop characteristics, and functions to change active and reactive power and voltage set-points. There are physical and virtual relays with over-current, over-frequency, and under-frequency protection functions integrated in different locations in the test feeder.

\subsubsection{\helics-\ns}
For our platform, as seen in Figure~\ref{fig:environment}, we use \helics as a interconnecting bridge between \gridlabd and \ns. At the start of a simulation run, the \helics-\ns node and \gridlabd connect to the \helics broker as federates and are ready to send/receive data from/to each other. Each \ns node is assigned a \helics endpoint that is used to communicate with the \helics broker. The \helics broker serves as the timekeeper for the simulation. When the \ns node receives a data request, it pings the \gridlabd tool for the updated values for the registered points. Similarly, when the \ns node receives a control command signal, it converts the command signal to \gridlabd setpoint change request to the \gridlabd tool via \helics broker. Once the request is fulfilled, the broker advances the simulation to the next timestep. This continues until the end of the simulation.

\subsubsection{\dnp-\ns}
We added \dnp protocol into \ns to simulate realistic grid communication scenarios between utility control center and the distributed grid. The \dnp-\ns module requires a configuration file consisting of all the measurement data expected from \gridlabd simulation. The data points are \gridlabd specific names of measurement data or set-points such as active power, voltage settings, ON/OFF states of switches etc. configured as analog or binary points. There are other types of points, but we focus on analog and binary points for the scope of this paper. To generate distribution grid scenarios, \ns node with \helics-\ns and \dnp-\ns modules enabled is needed. At the start of a simulation run, \gridlabd measurement points get registered with \gridlabd through \helics broker. \helics-\ns is responsible for retrieving measurements from \gridlabd tool and sending set-points to modify the DER settings. The \dnp-\ns module is responsible for converting raw \gridlabd measurement values into \dnp protocol format and also translating \dnp control commands into \gridlabd specific set-point instructions.

\begin{algorithm}
\caption{Packet interception algorithm at the stack. \emph{ContToSub} shows the information flow from control center to the substation. \emph{SubToCont} shows the information flow from the substation to the control center.}\label{alg:cap}
\begin{algorithmic}[1]
\State $victimAddr \gets$ Address of the victim node
\State $IsDestination \gets False$
\Function{CapturePacket}{p, addr}
    \If{$addr = victimAddr$}
        \State $IsDestination \gets True$
    \EndIf
\EndFunction
\Function{ContToSub}{AppHeader ActionID, UserData data, AppSeqNum seq}
    \State packet $\gets$ initiateReq(ActionID);
    \State transmit(packet, data, seq)
\EndFunction
\Function{SubToCont}{UserData data, $map<string,float> analog\_points$, $map<string,uint\_t> bin\_points$, AppSeqNum seq}
    \For{k in $analog\_points$}
        \State packet $\gets k$ 
    \EndFor
    \For{k in $bin\_points$}
        \State packet $\gets k$ 
    \EndFor
    \State transmit(packet, data, seq)
\EndFunction
\Function{SendNewPacket}{DestID, SrcID}
\If{IsDeatination}
    \State UserData $data.dest \gets DestId$
    \State UserData $data.src \gets SrcId$
    \State AddSeqNum seq $\gets$ current sequence number
    \State $map<string,float> analog\_points \gets$ updated analog points
    \State $map<string,uint16\_t> bin\_points \gets$ updated binary points
\EndIf
\EndFunction
 
\end{algorithmic}
\end{algorithm}

\subsection{Configurations}

\subsubsection{Network Topologies} Our tool takes a JSON file containing node to node connections, gives it to \ns, and uses it to build a topology. Normally, to make sure that the topology is valid, a user can use a topology generator such as \ns topology generator~\cite{halder2018ns3tcg}, but doing so requires prior knowledge of operating \ns. We simplify this process by reducing the knowledge needed to operate \ns to a configuration file, thereby removing the need for additional programs outside the ones already installed in our \docker container.

The configuration file allows control over the jitter of a node, the connection type between two nodes, and the topology of the network. For example, a user can create a configuration that generates a topology where there are two groups of nodes: The first group of nodes is connected using Carrier Sense Multiple Access (CSMA) following a mesh topology structure, and the second group is connected using a point to point connection using a star topology. Then both of the groups can be connected over \wifi so that data can be sent between each cluster. This example is one of many configurations that can be created.

\subsubsection{Cyber-Attack} We use JSON files to setup the attack in our setup. The configuration file takes in the start and end time of the attack, the number of attackers in the network and attack specific values. The attack-specific values include the parameters of the victim device that are being attacked, such as active or reactive power settings. It also includes the attack value that the attacker uses to update the value of the device parameter. Finally, the user can also select the attack scenario that they want to simulate on the network.

\subsection{Attack generation method}
To generate attacks, we use the internet module and the \dnp module to intercept a packet and update it to contain new data. If the destination address is found to the be victim address, then the attacker simulates the packet arriving to its original destination. Then, through the \dnp protocol, the attacker simulates sending a new packet containing the updated information or command.  The $CapturePacket$ function is set in the ipv4 l3 protocol module. The rest of the functions are developed through the \dnp module. In Algorithm~\ref{alg:cap}, the packet can be intercepted using the internet stack before it is sent to the target node(s).

\begin{algorithm}
\caption{\dnp packet update. }\label{alg:att}
\begin{algorithmic}[1]
 \State $AttackValues \gets$ Updated values the attacker uses
 \State $PointIDs \gets$ Point IDs that are attacked
 \State $NodeIDs \gets$ Node IDs that are attacked 
 \Function{GetIndex}{NodeID+PointID}
   \For{k in $analog\_points$}
       \If{$k = NodeID+PointID$}
           \State \Return $index$ of k
       \EndIf
   \EndFor
   \For{k in $bin\_points$}
       \If{$k = NodeID+PointID$}
           \State \Return $index$ of k
       \EndIf
   \EndFor
 \EndFunction
 \Function{SendAction}{AttackValues,PointIDs, NodeIDs}
    \For{k in PointIDs and c in NodeIDs}
        \State $index \gets GetIndex(c+k)$
        \State UpdatePointValue(index)
    \EndFor
    \State ForwardPollRequest()
 \EndFunction
\end{algorithmic}
\end{algorithm}

\subsubsection{Our attacker}
Our attacker is a  \MIM attacker who has access to a network node using a \dnp application within \ns. The attacker uses a modified ipv4 l3 protocol to identify if the message that is intercepted is heading to the ipv4 address of the victim. Once the victim address is identified, the attacker captures the messages, and sends an updated message with new data in place of the old. Before the message is sent, the source address is updated to match the address of the original sender. Thus, once the receiver gets the message, they will act upon it as if it was from a legitimate source. The attack can be used as a standalone attack where the goal is to spread misinformation; or it can be used as a part of a larger attack where the goal would be not only to spread misinformation but also to take control over a node/section of the network. 

For this attack scenario, the attacker intercepts traffic going from the control center to the substation or the microgrid or from the substation to the control center. When intercepting the packet going from the control center to the substation, the attacker modifies the packet to not only conduct the normal action that was requested from the control center but to also conduct an action chosen by the attacker. For example, if the control center sends out a poll request to the individual nodes in the network, the attacker can intercept that message and send an action to the recipient to trip a relay and modify the resulting data that is returned by the substation to hide the changed values. By intercepting the data going from the substation to the control center, the attacker can send fake data to the control center as an attempt to trick the control center into thinking that everything is fine at the substation level.

\section{Experimental Setup}

This section describes generation of three types of attack scenarios on a distribution grid. We also demonstrate how to use our tool to modify where the attacker is and how the attack impact varies depending on the location of the attacker in the network. For all of the three attack scenarios described bellow, the attacker is located between the control center and the substation.

\begin{itemize}
\item In the first attack, the attacker intercepts the measurement data flowing from substation to control center and modifies the reactive power setpoints/reference (Qref) of Inverter 42 (grid-following inverter) situated in Microgrid 1. 

\item In the second attack, the attacker intercepts the \dnp command flowing from control center to the substation and modifies the active power setpoints (Pref) value for both grid-following and grid-forming inverters -Inverter 42 and Inverter 51 of Microgrid 1. Additionally, during both of these attacks the switches connecting the microgrids are tripped to island the microgrids from each other and the grid. 

\item In the third attack, the attacker intercepts the \dnp command flowing from control center to the substation and islands the microgrids from each other and the grid. We conduct that attack to examine the impact of an islanding attack on the frequency measured on Microgrid 3. Islanding happens when microgrids are disconnected from one another, making the microgrids dependent on their individual power sources. Finally, once we find a set of parameters that can be used to counter the effect of islanding on the frequency, we trip an internal virtual relay to cause extra load shedding. 

\item In the fourth attack, the attacker is in the same location as with attack 2 and 3. We conduct the two previous attacks on a ring topology. The attacks in the previous scenarios were conducted in a star topology. The goal of this attack is to compare the effect of both attacks on the analog and binary points that are sent from the substation to the control center on a different topology. 

\end{itemize}

\subsection{Data format and setup}

The experiment setup has a control center that is responsible for monitoring the distribution grid of an area. The grid consists of the IEEE 123 test feeders with three microgrids. The grid dynamics is simulated using \gridlabd tool. The microgrids can be islanded by tripping the relay/switches connecting them to each other and the grid. Each microgrid has a substation which acts as the remote terminal units (RTUs) that aggregate data from and disseminate control signals coming from the control center to various DERs in the microgrid. The control center and substation are setup as \ns nodes and communicate over virtual \ns network. The substation nodes have \helics-\ns to integrate with \gridlabd. The control center and substations have \dnp-\ns module installed to provision DNP3 based grid communication. The control center makes a periodic \dnp polling request (every 4 seconds) to each of the substations for latest measurement values from the microgrid. Each of the substations respond back with data collected from \gridlabd simulation for the respective microgrid. The control center can also send \dnp control command to a substation to control a particular set-point of DER in the microgrid. The substation acts on the control command by sending write instruction to the \gridlabd simulation.

During the configuration setup, all the \gridlabd specific measurement data such as active power, reactive power, ON/OFF state of switches are pre-configured as \dnp analog and binary points in a configuration file. During simulation start-up, these data points are ingested by \ns-\helics module at the substation node and registered with \gridlabd. The raw measurements are collected and converted into DNP3 protocol format. The packaged data is then sent to the control center for processing over the network. 

\begin{figure*}[!htb] 
  \centering
  \includegraphics[width=1\linewidth]{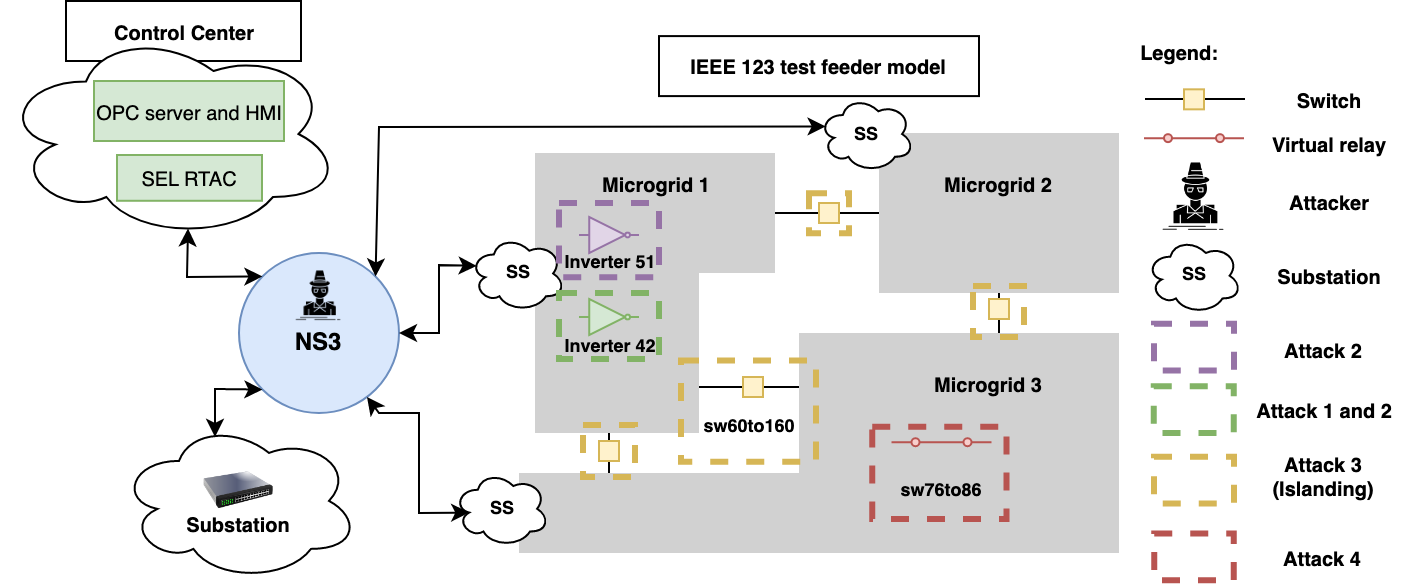}
  \caption{Microgrid setup for experimentation, using the IEEE feeder model as described by Ashok~\emph{et al.}~\cite{ashok2021high}. We use this setup to run the cyber attacks and collect data on how the attacks impact the performance of the power grid. The attack conducts a \MIM attack on two inverters in Microgrid 1, the switches connecting the Microgrids and the relay in Microgrid 3. Substations can be responsible for power distribution over multiple Microgrids as well as act as aggregate points.}
  \label{fig:microgrid}
\end{figure*}

Figure~\ref{fig:microgrid} shows the system architecture with the design of our power grid, and the grid consists of the IEEE 123 test feeders with three microgrids. The microgrids can be islanded by tripping the relay/switches connecting them to each other and the grid.
The control center and the substation are also connected to the router where \ns is loaded to allow different topologies during experimentation.

\subsection{Topologies tested}

\begin{figure*}[!htb] 
  \centering
  \includegraphics[width=0.8\linewidth]{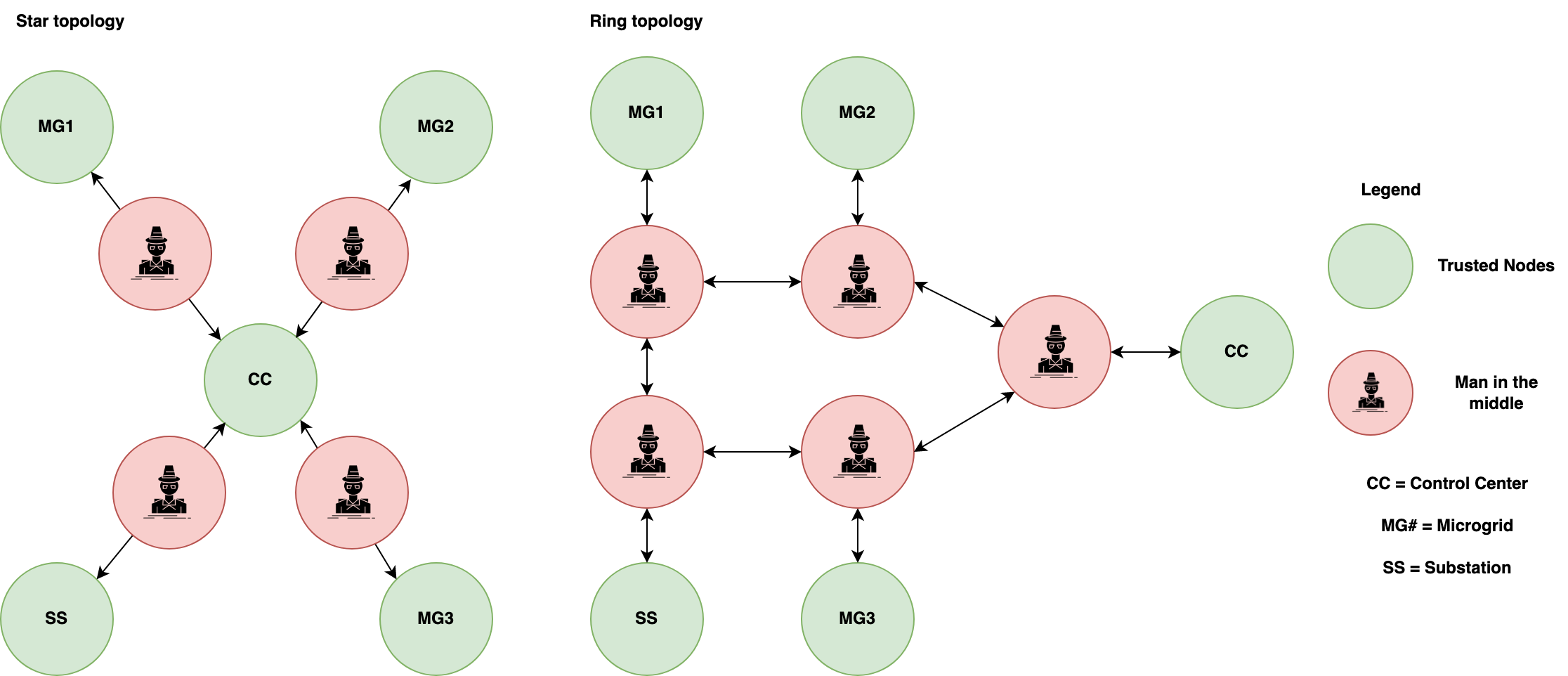}
  \caption{A simplified presentation of the topologies used in our experiments. The Ring topology was defined using our topology configuration file while the star topology was defined directly in our code as the default topology in case the user does not have a specific topology to use.}
  \label{fig:topology}
\end{figure*}

We run our attacks on two topologies: a default star topology that has the control center at the center of the network and a ring topology built using our topology configuration file. Figure~\ref{fig:topology} illustrates both of the topologies.

\subsection{Attack scenarios}
We place an attacker located between the control center and the microgrid/substation. The attacker must stay unnoticed since it does not have authority to be on the network. Using a modified Internet Stack, the attacker reads the bytes in the packet and, if it finds a certain packet that matches the requirement for the attack, executes an interception. The attacker then modifies the intercepted packet data to communicate false information to the control center. This attack can be also used in conjunction with a command injection. Using both, an attacker can perform changes to the settings of a node while hiding it from the control center. 

\subsubsection{Attack 1: Data Modification}
In this scenario, the attacker intercepts a poll request response going from the substation to the control center. The attacker then modifies the reactive power setpoints/reference (Qref) of Inverter 42 (grid-following inverter) situated in Microgrid 1 to trick the control center to believing that the inverter is in a different state then its current state. 

\subsubsection{Attack 2: Inverter active/reactive power setpoints modifications}

\begin{table*}
\begin{center}
\begin{tabular}{ |p{2cm}|p{2cm}|p{2cm}|p{2cm}|p{2cm}|p{2cm}|p{2cm}|  }
 \hline
 \multicolumn{7}{|c|}{Inverter Description} \\
 \hline
 Inverter ID & location & type & Rated & Pref & Qref & attacked values\\
 \hline
 Inverter 51  & MG1 & grid forming   & 400 kW & 210 kW & 0 VAR & Pref (110 kW)\\
 Inverter 42  & MG1 & grid following & 600 kW & 450 kW & 0 VAR & Pref (350 kW), Qref ($-50$~kVAR)\\
 Inverter 101 & MG2 & grid following & 180 kW & 126 kW & 0 VAR & NA\\
 Inverter 105 & MG2 & grid forming   & 600 kW & 300 kW & 0 VAR & NA\\
 Inverter 76  & MG3 & grid following & 120 kW & 84 kW  & 0 VAR & NA\\
 Inverter 80  & MG3 & grid forming   & 100 kW & 70 kW  & 0 VAR & NA\\
 \hline
\end{tabular}
\caption{Available Inverters in our current simulation with default and attack values.}
\label{Inve}
\end{center}
\end{table*}

In this scenario, the microgrids are islanded from other microgrids and the grid. Then, a \MIM (MITM) attack changes the setpoint of an inverter to introduce voltage issues. The attack happens at approximately two minutes into simulation. As seen in Table~\ref{Inve}, Inverter 42 has its Qref setpoint changed from 0~Var (default) to $-50$~kVar (attack value). In the second scenario, the microgrids are also islanded. Then, inverters are attacked consistently to cause voltage stability issues: Inverter 42 has its Pref value randomly toggled between 450~kW (default) and 350~kW (attack value) and Inverter 51 has its Pref value randomly toggled between 210~kW (default) and 110~kW (attack value). The attack starts at approximately two minutes into simulation.

This scenario uses packet capture (PCAP) files that are generated by \ns to identify the attack and quantify its impact on the resulting network. \emph{PCAP} files are used as ways to visualize network traffic. \ns can generate them using the point to point helper class in the point to point module and can be read using wireshark. 

\subsubsection{Attack 3: Tripping relays using command injection}

\begin{table*}
\begin{center}
\begin{tabular}{ |p{2cm}|p{2cm}|p{2cm}|p{2cm}|p{2cm}|  }
 \hline
 \multicolumn{5}{|c|}{Relay Description} \\
 \hline
 Relay/switch ID & location & type/use & default value & attacked value\\
 \hline
 sw60to160           & grid-MG2 & Breaker/Switch & close & open\\
 sw18to135           & grid-MG1 & Breaker/Switch & close & open\\
 sw97to197           & MG2-MG3  & Breaker/Switch & close & open\\
 sw54to94            & grid-MG3 & Breaker/Switch & close & open\\
 sw151to300          & MG1-MG2  & Breaker/Switch & close & open\\
 sw76to86            & MG3      & Virtual Relay  & close & open\\
 \hline
\end{tabular}
\caption{List of relays connecting the microgrids to one another and to the grid. These relays are used in our experiment to island the microgrids from one another. The Virtual Relay is used as a load disconnect to cause additional Under-Frequency-Load-Shedding.}
\label{Tab:Rel}
\end{center}
\end{table*}

In this scenario, a command injection attack causes islanding of microgrids, as shown in Table~\ref{Tab:Rel}. In some cases, power generation is enough to sustain the loads running on the microgrid, while in others (i.e. Microgrid 3) limited power generation increases the Under-Frequency-Load-Shedding (UFLS). UFLS occurs due to insufficient generation on the microgrids to supply the load. The sw60to160 relay was tripped via command injection at approximately two minutes into data capture. In another version of this attack, power dispatch has been adjusted to minimize UFLS. Finally in a third version of this attack, the sw60to160 relay is tripped and the config file of the sw76to86 virtual relay was modified so that additional UFLS would occur.

For this experiment, we use frequency measurement to compare the three parts. Frequency has been used to identify any load mismatches that may be happening in a power system~\cite{lobos1997real,yang2001precise}. When a cyberattack occurs, its impact on the point value of nodes in a power system causes the frequency to shift away from its normal, pre-attack, value. 

\subsection{Attack 2 and 3 on a Ring topology}
For this scenario, we run attack scenario 2 and 3 on a ring topology. This scenario serves as an example of how to use the dynamic topology configuration function of our tool. We compare the resulting effect of the attack on the collected analog and binary points collected over the grid.

\section{Results}
During experimentation, we collect information described in the previous section to demonstrate the operation of our tool. We identified how changing certain parameters, such as the active power of an inverter (Pref) value changes, can cause a grid following and grid forming inverter to have different output voltage behavior. Additionally, when looking at the microgrid's frequency change during an attack, by changing the generator's nominal and output power, the frequency was able to return to its value before the attack started. 

\subsection{Attack 1: Data Modification}

We start the experiment by examining the response that is sent from the substation to the control center in response to a poll request. The substation's response is sent in three segments of 274 bytes. This response message contains points, each point representing a value associated with a node. These node points, and their values, represent the behavior of that node at the time the packet was sent, for example, the output voltage of generator 1. Each of these points are separated by a flag, and the flag represents whether or not that point was set correctly. 

\begin{figure}[!htb] 
  \centering
  \includegraphics[width=1\linewidth]{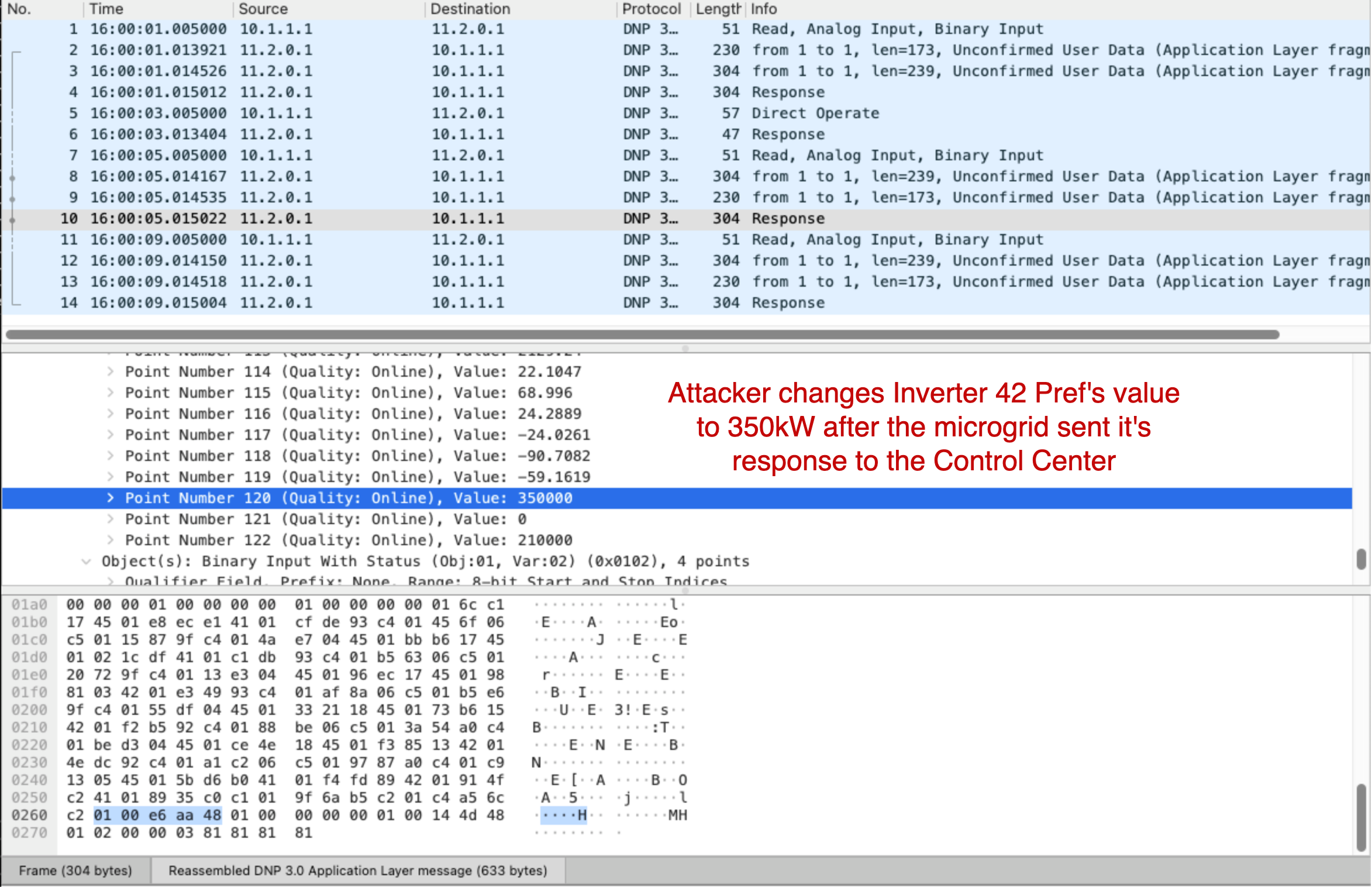}
  \caption{The attacker modifies the response from the substation to the control center to make the control center believe that the Pref value of inverter 42 is set to 350~kW instead of 450~kW, its default value.}
  \label{fig:PCAP-exp1}
\end{figure}

An attacker who has access to these response packets can use a history of these packets to build an understanding of what nodes constitute the microgrid. For this experiment, we intercept some of these response packets, and change several points inside them, as seen in Figure~\ref{fig:PCAP-exp1}. After the bytes are updated by the attacker, the attacker sends out the updated packet with the new values back into the network. Our tool can generate PCAP files though the \ns simulator, which can be used to visualize this attack. 

\subsection{Attack 2: Inverter active/reactive power setpoints modifications}
\begin{figure}[!htb] 
  \centering
  \includegraphics[width=1\linewidth]{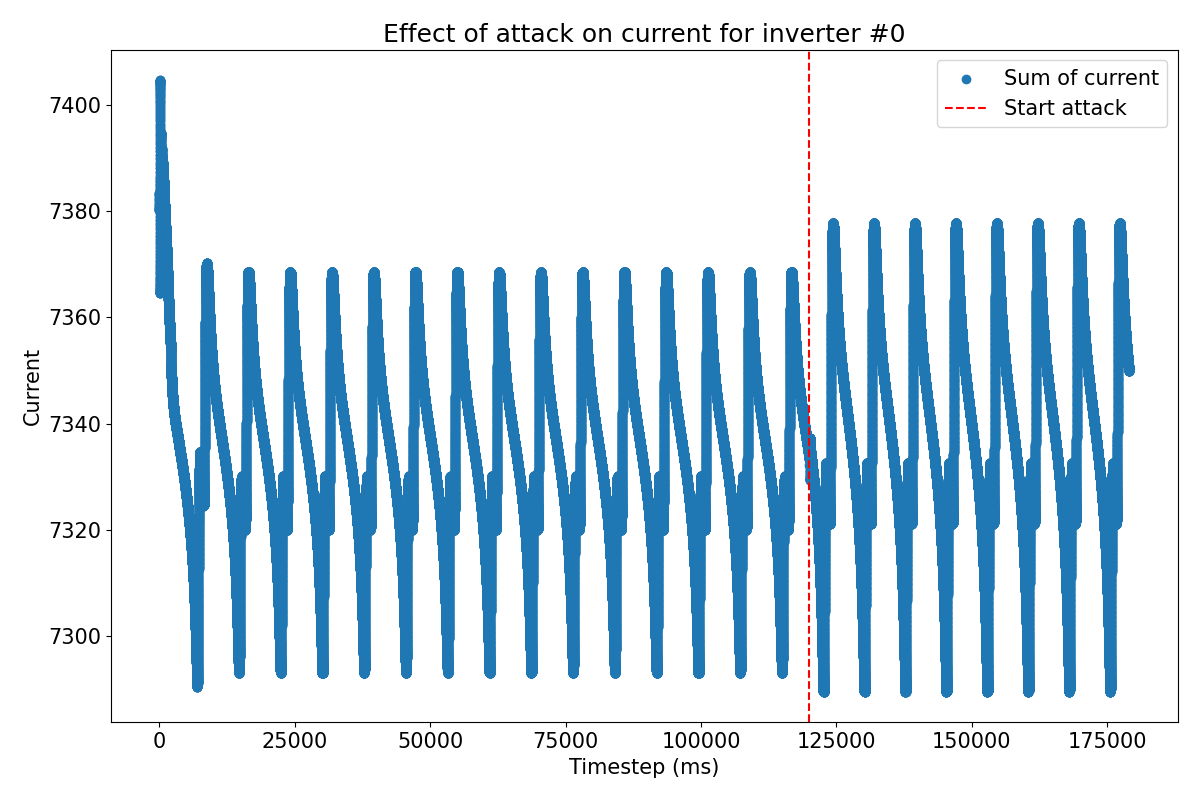}
  \caption{Current change once the attack starts at load 42, which is the load connected to inverter 42. The attack starts 2 minutes in the simulation and causes a shift in the current fluctuation of the load, where the current waves became bigger compared to before the attack.}
  \label{fig:current}
\end{figure}

\begin{figure}[!htb] 
  \centering
  \includegraphics[width=1\linewidth]{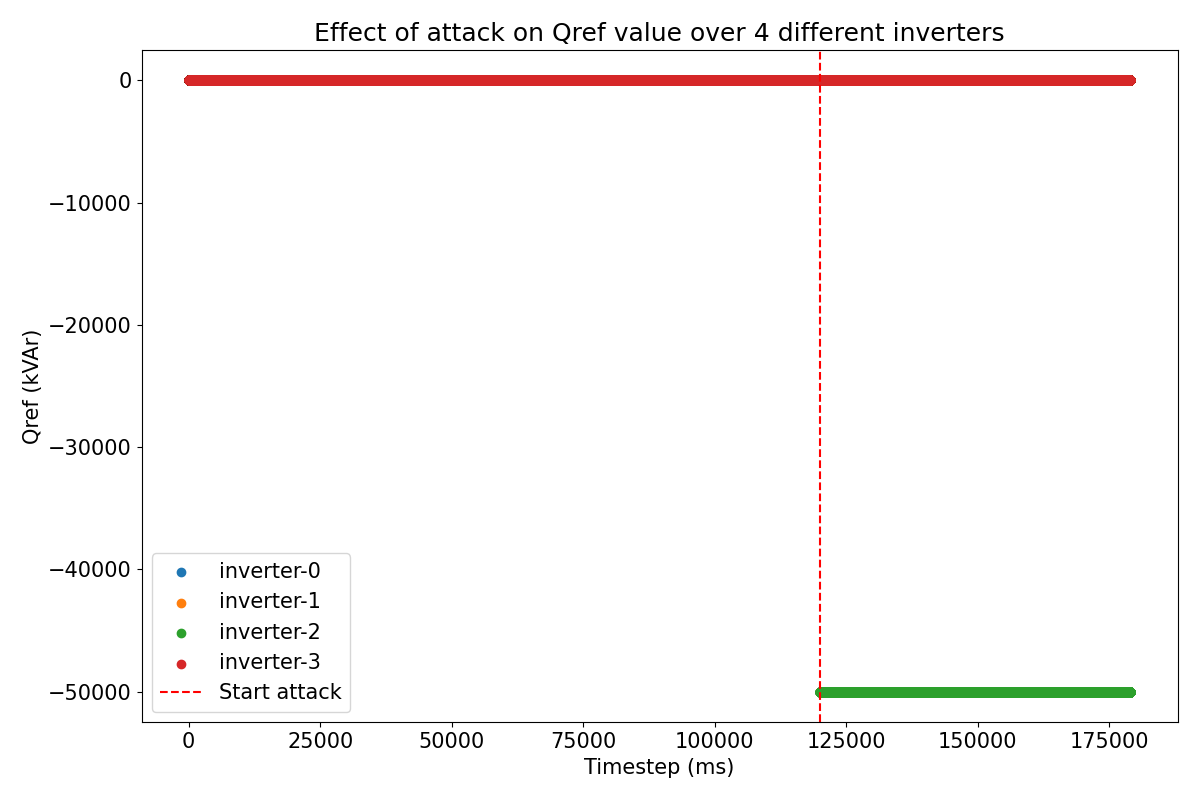}
  \caption{The Qref of inverter 42 (inverter-2) is dropped to $-50$~kVAR from 0~kVAR (the default value) after 2 minutes in the simulation. The rest of the inverters are not attacked.}
  \label{fig:Qref}
\end{figure}

The attacker starts by reducing the Qref value of inverter 42 two minutes into the simulation, as seen in Figure~\ref{fig:Qref}. This causes a slight decrease in current, as seen in Figure~\ref{fig:current} where the current normally averages between 7359~W to 7346~W. This very slight increases in maximum and decreases in minimum, resulting in larger waves, in current may go unnoticed if there is no existing knowledge that a change similar to this observed change signifies an adversary on the network. Reactive power (Qref) maintains voltage levels that are needed for system stability. Therefore, it makes sense that by reducing the Qref value, the current value changes over time. 

Our tool can visualize these changes to make informed decisions on how they can tackle such attacks. A user can use the datasets generated by our tool to train models to identify and counter attackers based on the impact that is viewed at the node level. In this case, if an inverter's current suddenly drops, the traffic can be rerouted away from that inverter and microgrid to isolate that section of the network. Additionally, the network can send a signal that a section of it is under attack so that some countermeasure can be applied to eliminate the threat. 

\begin{figure}[!htb] 
  \centering
  \includegraphics[width=1\linewidth]{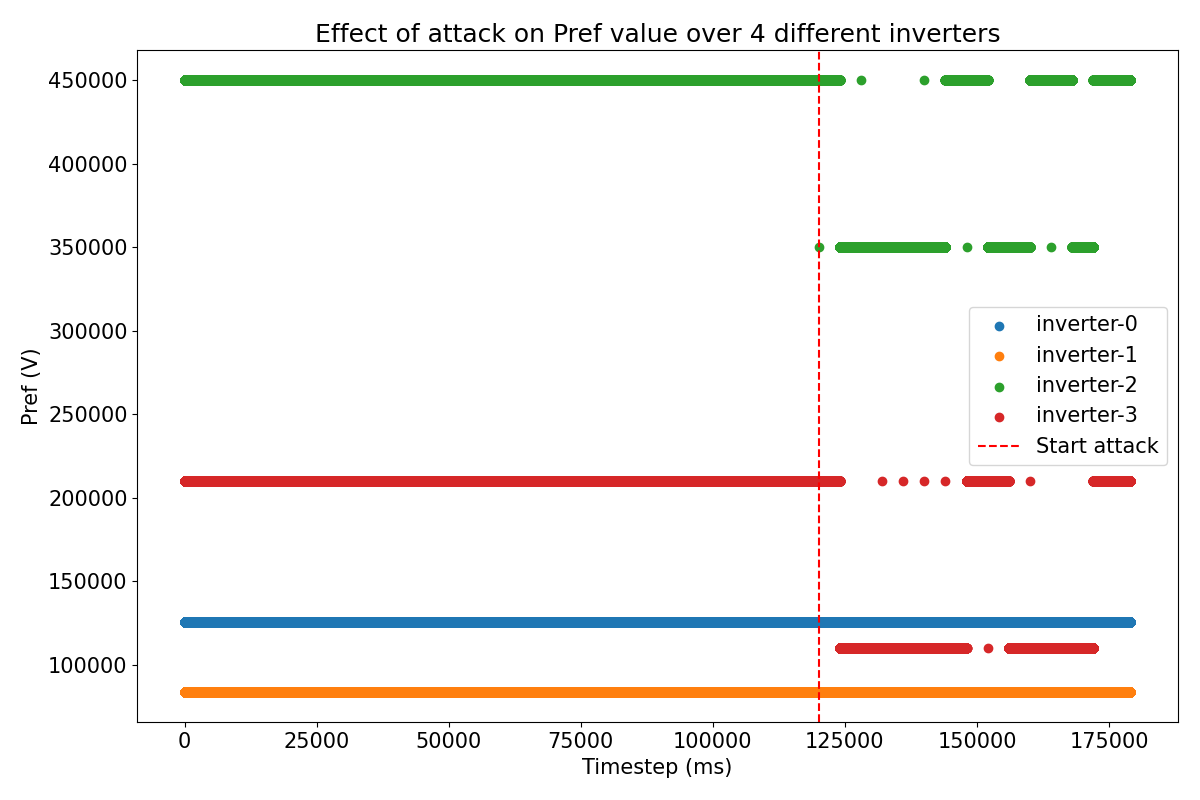}
  \caption{The Pref of Inverter 42 (inverter-2) is dropped to 350~kW from 450~kW (the default value) and the Pref value of Inverter 51 (inverter-3) is dropped from 210~kW to 110~kW after two minutes in the simulation. This is an example of the Pref values being randomly fluctuating between the default and attack values.}
  \label{fig:Pref}
\end{figure}

Our tool can also be used to simulate a multi-node attack, as demonstrated in the second part of this attack. Here, the attacker randomly varies the Pref value of two inverters on Microgrid 1 as seen in Figure~\ref{fig:Pref}.

\begin{figure}[!htb] 
  \centering
  \includegraphics[width=1\linewidth]{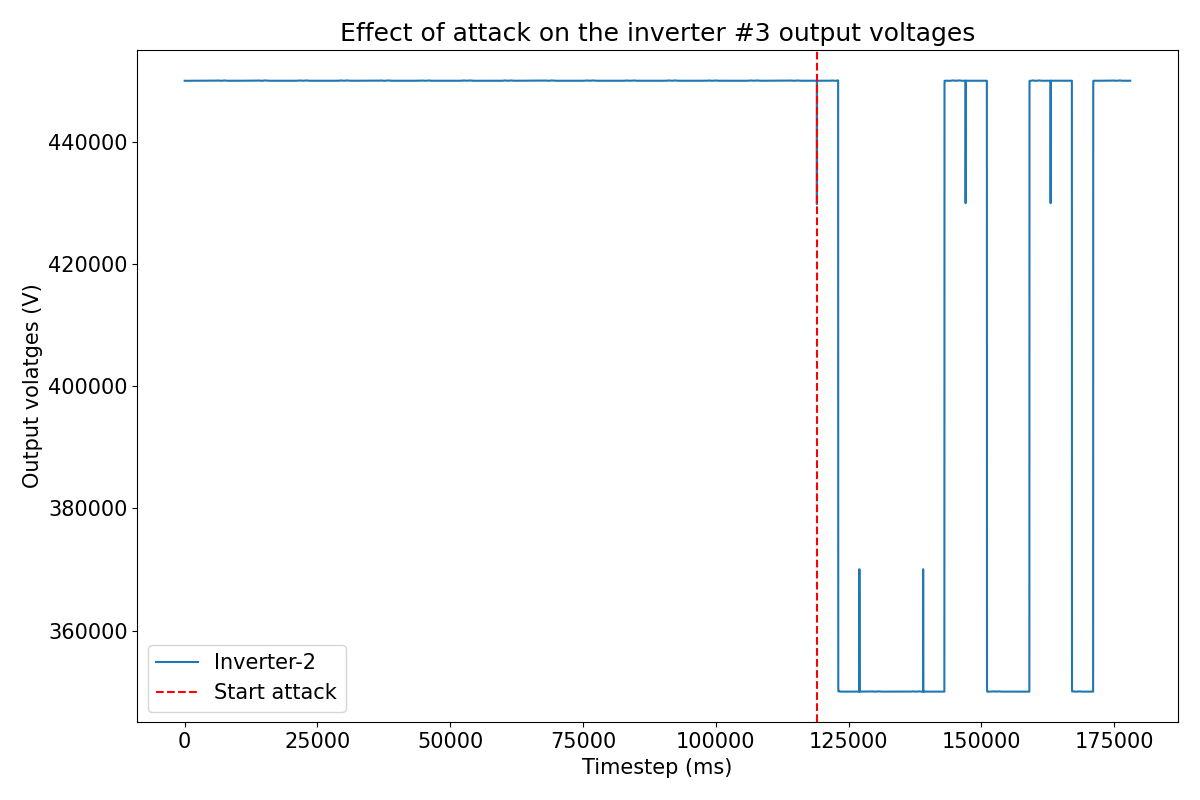}
  \caption{Effect of attack on inverter 42's, a grid following inverter, output voltage.The attacker dynamically fluctuates the Pref value of two distinct inverters to affect the resulting output voltage and current of the microgrid. In this scenario the microgrids are islanded from both each other and the grid.}
  \label{fig:VAout}
\end{figure}

Figure~\ref{fig:VAout} shows a drastic decrease in output voltage of Inverter 42 during the attack where the attacker randomly increases and decreases the Pref value. As seen in Figure~\ref{fig:Pref}, the variation in Pref value matches the variation in output voltage for inverter 42 (inverter-2). The output voltage drops to the attack value of 350~kW when the attacker sets the Pref value to 350~kW. By identifying these drop in voltages, a user of our tool can create counter measures to such attacks by potentially resetting the inverter if the voltage fluctuates erratically.  

\begin{figure}[!htb] 
  \centering
  \includegraphics[width=1\linewidth]{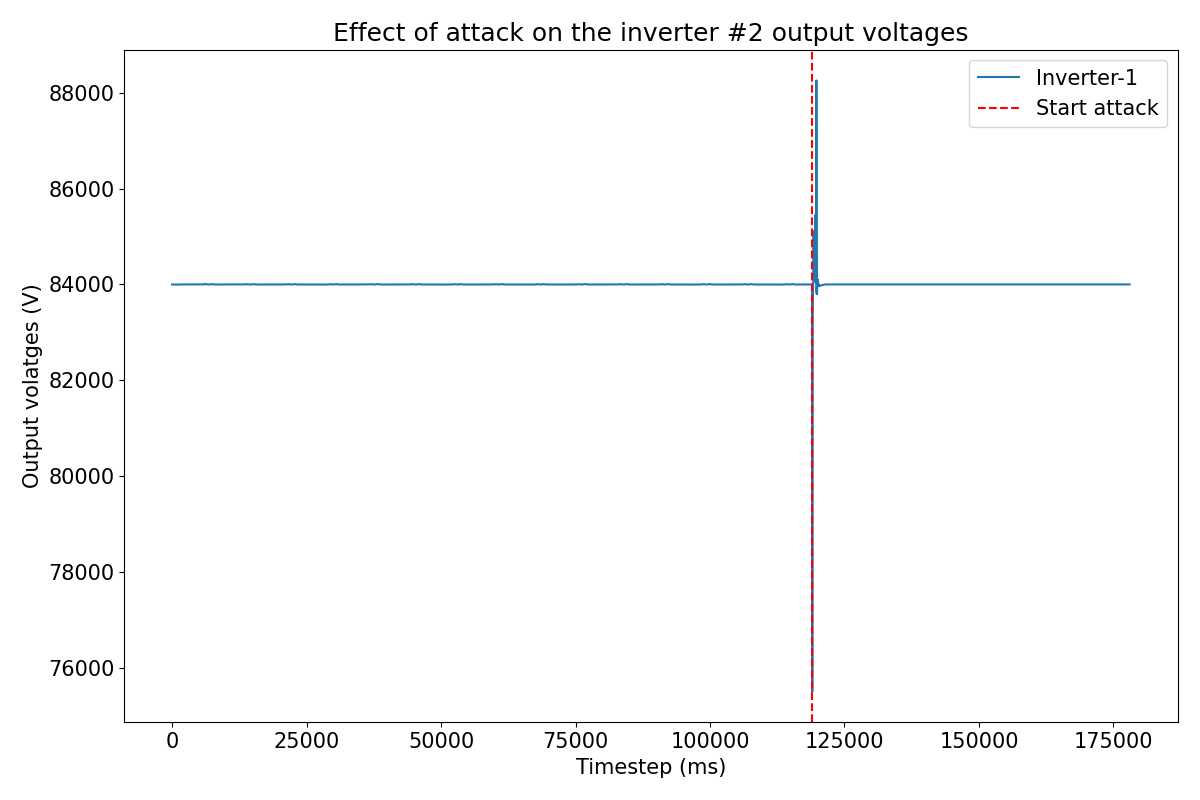}
  \caption{Impact of attack on inverter 42 and inverter 51 on an inverter's output voltage in a different microgrid}
  \label{fig:VAout_MG2}
\end{figure}

Figure~\ref{fig:VAout_MG2} shows how the attack happening on Microgrid one's inverter can affect the resulting output voltage in inverters in a seperate microgrid. Since Microgrid 1 is islanded from the rest of the Microgrids and the grid when the attack starts, we can observe an increase in power fluctuation when looking at the output voltage of the inverters.

\begin{figure}[!htb] 
  \centering
  \includegraphics[width=1\linewidth]{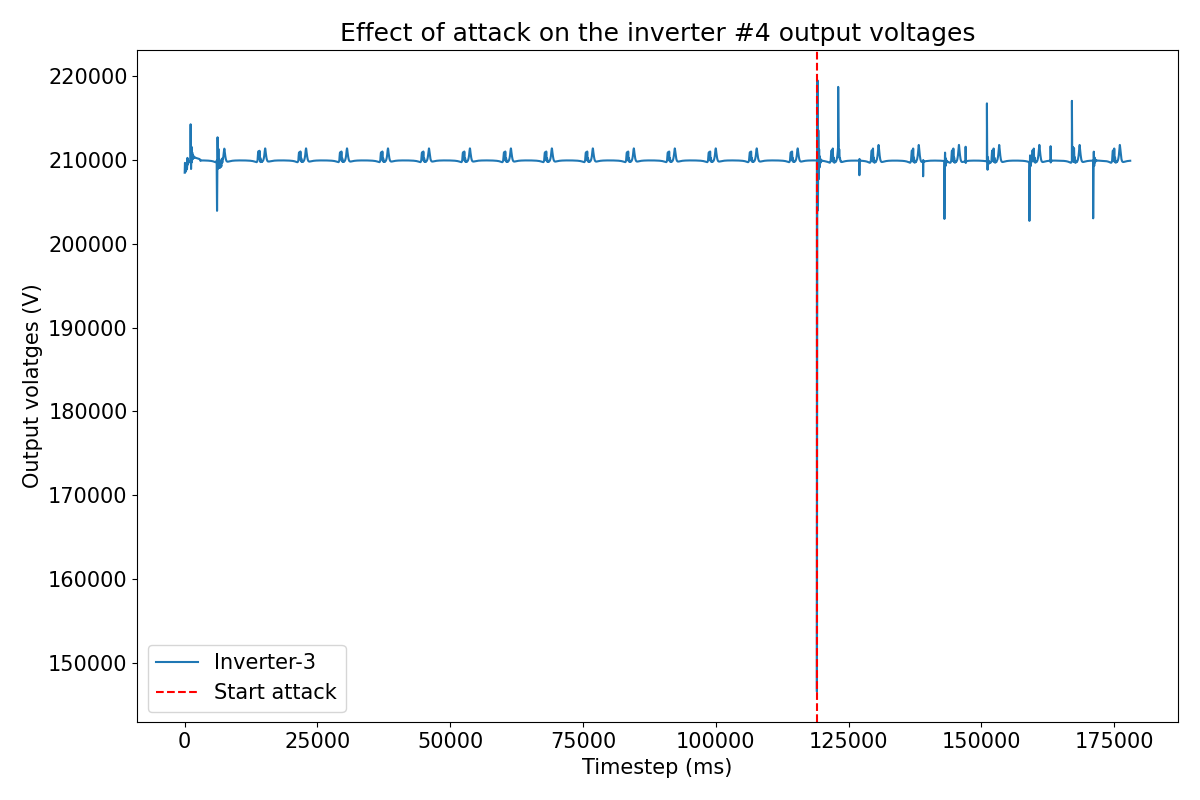}
  \caption{Impact of attack on inverter 51's, a grid forming inverter, output voltage}
  \label{fig:VAout_MG1}
\end{figure}

Figure~\ref{fig:VAout_MG1} shows the impact of the attack on a grid forming inverter. In this case, we can see a significant drop in the inverter's output voltage, but it is quickly brought back to its normal value after a few timesteps. The fluctuations observed in Figures~\ref{fig:VAout} and~\ref{fig:VAout_MG1} show that a grid forming inverter is less susceptible to attacks on the Pref value compared to a grid following inverter. 

\begin{figure}[!htb] 
  \centering
  \includegraphics[width=1\linewidth]{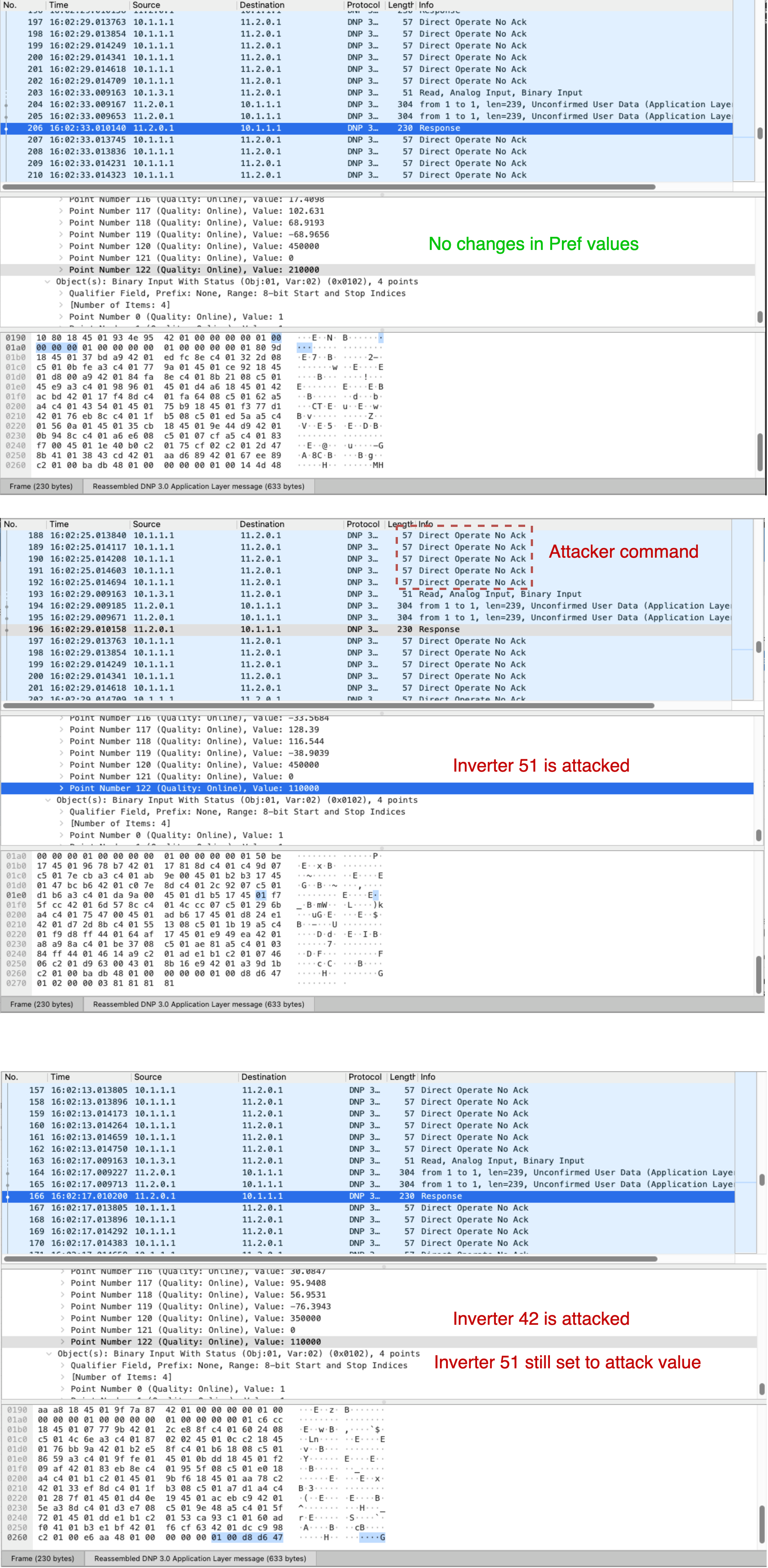}
  \caption{Visualization of attack through PCAP file.}
  \label{fig:PCAP}
\end{figure}

We can also visualize the attacks with PCAP files. Figure~\ref{fig:PCAP} shows how the Pref value changes over time. Using PCAP, we can see the cycle of commands that is sent by the attacker to the inverter. We can also see that the only points that get modified are the Pref value of both of the inverters. Interestingly, the size of the packet that is received as a response from the microgrids varies between 230 Bytes and 304 Bytes as the Pref fluctuates. This is another indication that can be used by a user to identify an attacker on the system.  

\subsection{Attack 3: Tripping relays using command injection}

\begin{figure}[!htb] 
  \centering
  \includegraphics[width=1\linewidth]{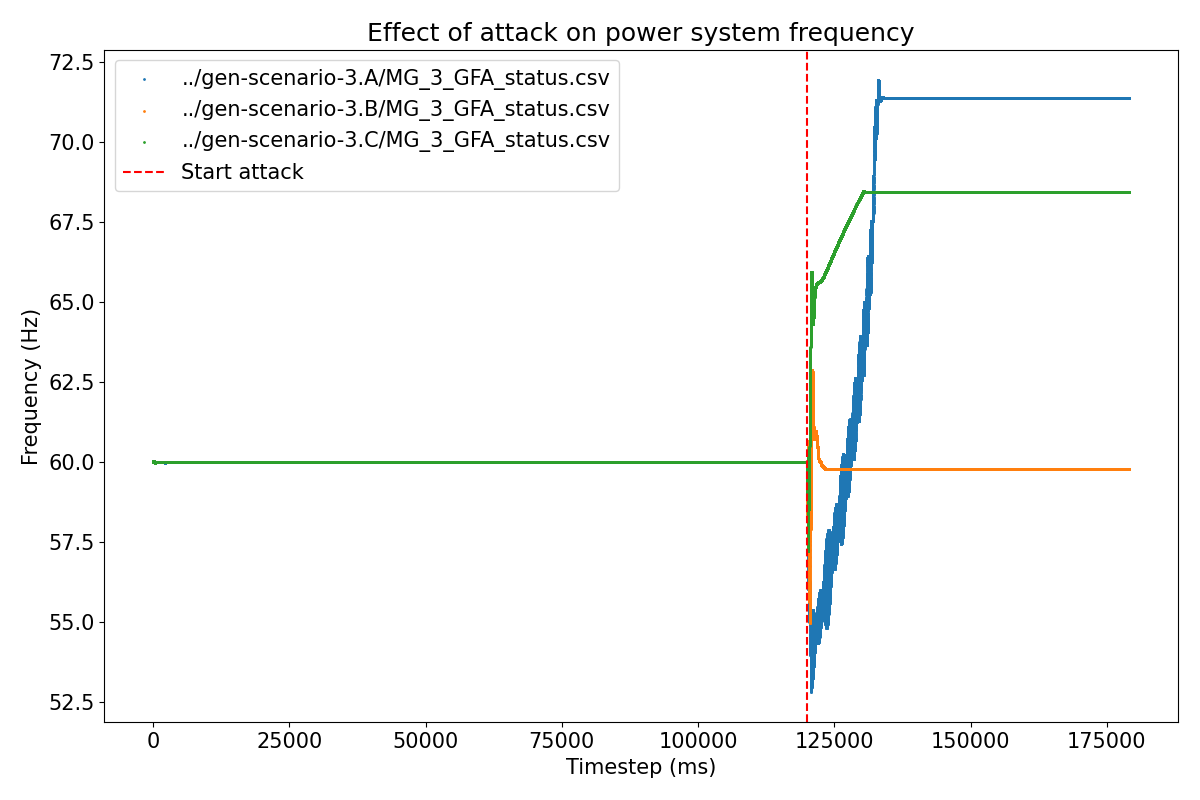}
  \caption{Frequency changes depending on when the attack starts and what changes, if any, are done to reduce the frequency deviation.}
  \label{fig:frequency}
\end{figure}

In this scenario, we are trip the relays connecting the microgrids to one another, as seen in Table~\ref{Tab:Rel}. We look at the frequency measurement of the microgrids to identify how resilient it is to attacks. Tripping the relays can cause the microgrids to become islanded from one another and the grid. Notably in this attack, microgrid 3, as shown in Table~\ref{Tab:Cap}, is the only microgrid that has capacitors and it contains one large capacitor connected to multiple phases and three smaller capacitors connected to individual phases. In Figure~\ref{fig:frequency}, we can see that the frequency of the attacked microgrid increases to around 71~Hz during scenario A, the scenario where the generator and capacitor use the default value for our tool, as seen in Table~\ref{Tab:Gen}. Out of all the Microgrids, Microgrid 3 has the lowest power generation, causing it to struggle if Microgrid 3 is islanded, such as the result of tripping the relays in this attack. 

\begin{table*}
\begin{center}
\begin{tabular}{ |p{3cm}|p{3cm}|p{3cm}|p{5cm}|  }
 \hline
 \multicolumn{4}{|c|}{Microgrid 3 Capacitor Default values} \\
 \hline
 ID & phases & Nominal power & Capacitor size information\\
 \hline
 cap83 & A, B, C & 2401.7771~V  & A=200~kVAr, B=200~kVAr, C=200~kVAr\\
 cap88 & A & 2401.7771~V  & A=50kVAr\\
 cap90 & B & 2401.7771~V & B=50kVAr\\
 cap92 & C & 2401.7771~V & C=50kVAr\\
 \hline
\end{tabular}
\caption{Microgrid 3 is the only microgrid containing capacitors. The size information is in regard to the capacitor size that is connected to a specific phase. For example, in this table, A=200~kVar represents the size of the capacitor connected to phase A.}
\label{Tab:Cap}
\end{center}
\end{table*}

\begin{table*}
\begin{center}
\begin{tabular}{ |p{3cm}|p{3cm}|p{3cm}|p{3cm}|  }
 \hline
 \multicolumn{4}{|c|}{Synchronous Generator Default values} \\
 \hline
 ID & location &  rated power output & power delivered to interconnected nodes\\
 \hline
 Gen1 & MG1 & 10~MW  & 30~kW+3000~j\\
 Gen2 & MG1 & 1~MW   & 25~kW+8333~j\\
 Gen3 & MG3 & 450~kW & 50~kW+16667~j\\
 Gen4 & MG2 & 600~kW & 50~kW+16667~j\\
 \hline
\end{tabular}
\caption{\gridlabd simulates 2 types of generators (synchronous vs induction). The microgrids only use synchronous generators. Microgrid 3's generator is rated with the lowest power level out of the rest of the generators. It also delivers the most (tied with Gen4) power to the interconnected nodes.}
\label{Tab:Gen}
\end{center}
\end{table*}

Capacitors are a useful tool when regulating the frequency of a microgrid during an attack. When the simulation ran with a larger capacitor and a generator with a lower nominal power rating and lower power amount delivered to interconnected nodes, as seen in Table~\ref{Tab:Att}, the frequency returns to the normal pre-attack value after approximately two minutes. 

\begin{table*}
\begin{center}
\begin{tabular}{ |p{3cm}|p{3cm}|p{3cm}|p{5cm}|  }
 \hline
 \multicolumn{4}{|c|}{Updated Generator and Capacitor values} \\
 \hline
 ID & location &  rated/Nominal power & delivered power/capacitor size\\
 \hline
 cap83 & A, B, C & 2401.7771~V  & A=600~kVAr, B=600~kVAr, C=600~kVAr\\
 Gen3 & MG3 & 300~kW & 20~kW+16667~j\\
 \hline
\end{tabular}
\caption{All three phases are increased to 600~kVAr. Both the rated and delivered power for Generator 3 are reduced for this part of the experiment.}
\label{Tab:Att}
\end{center}
\end{table*}

Finally, when we trip the virtual relays between nodes 76 and 86, as seen in Figure~\ref{fig:microgrid}, while keeping the changes created in scenario B, we can see that the frequency does not return to the value before the attack started. Additionally, the frequency does not drop as much as during the scenario A attack. We also observe that after an initial climb, the frequency slightly drops before starting to climb at a slower rate. The frequency finally stabilizes at around 140 seconds.

\subsection{Attack 2 and 3 on a Ring topology}

Using our topology configuration file, we change the resulting topology from a star topology to a ring topology, as seen in Figure~\ref{fig:topology}. Each substation/microgrid/control center node is connected to a ring of intermediate nodes. The nodes on the ring are then used to conduct the \MIM attack. We conduct the same attack as in the previous section where we modify the Pref and Qref value of two different inverters located on the same microgrid. For this topology, similar to the previous topology, the nominal power was only lowered to 300~kW. Using our tool, we found out that to mitigate frequency changes when Microgrid three is islanded the same parameter values for the generator and the capacitor worked for both the ring and star topology.

\begin{figure}[!htb] 
  \centering
  \includegraphics[width=1\linewidth]{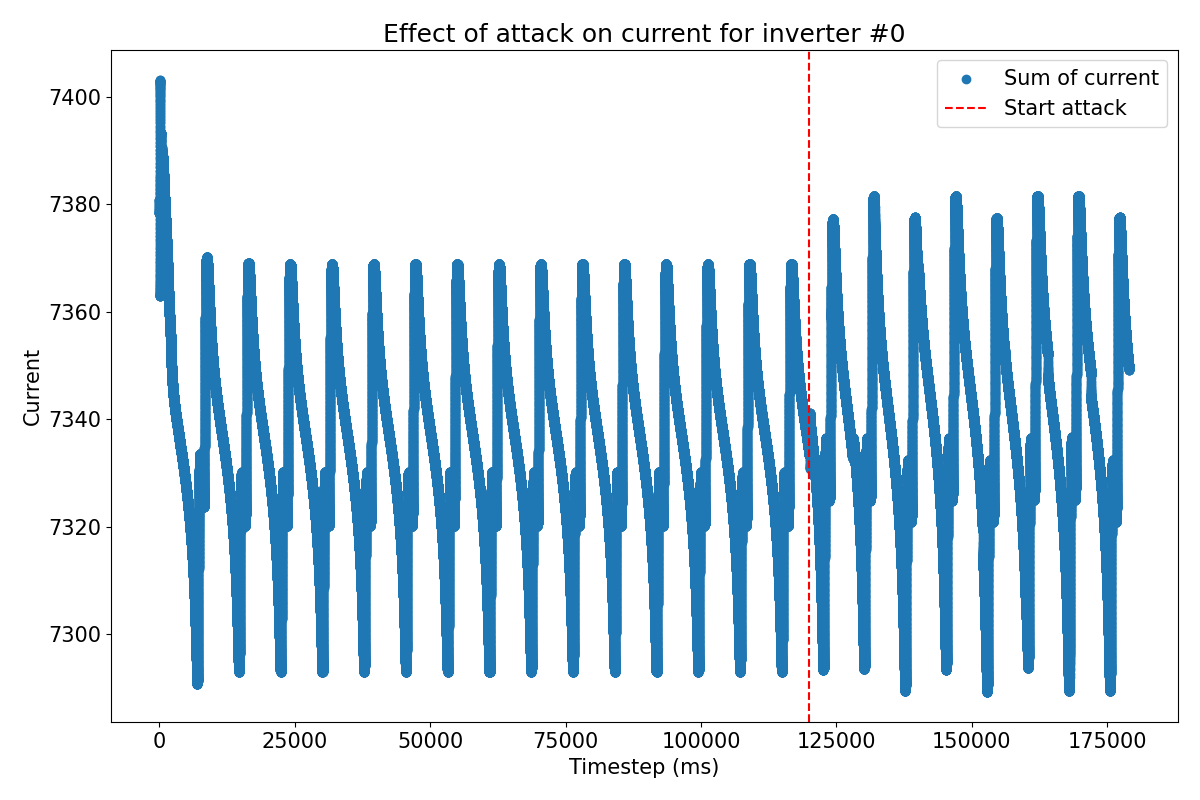}
  \caption{Current change once the attack starts at load 42, which is the load connected to Inverter 42 when the Pref value of both Inverter 42 and 51 are fluctuating. The attack starts 2 minutes into the simulation and causes a shift in the current values of the load.}
  \label{fig:current-ring}
\end{figure}

With regards to attack 2, where the Qref and Pref values are modified, we can observe that on this topology, when the Qref or Pref values are lowered, that the resulting current measured at the inverter fluctuates more similar to what was observed in a star topology. Figure~\ref{fig:current-ring} shows the resulting current measured at the connected load to inverter 42 in Microgrid 1 when the Pref value for both inverter 42 and 51 have their Pref value fluctuating over time. We can see that the waves have both a smaller minimum and a higher maximum similar to the current graph for the star topology. 

\section{Conclusion and future work}
We demonstrate the benefit of using a lightweight tool to model different security scenarios and their effect on grid nodes. We also described the design of our tool, and how it uses the IEEE model from \gridlabd to model grid components. Using our tool, we identified that there is differing behavior between grid forming and grid following inverters when the Pref and Qref values are changed by an attacker. Grid following inverters are more affected by any changes done to the Pref value, while grid forming inverters can restabilize the output voltage of the inverter down to its original value. We also identified parameter settings that enable the microgrid to recover the original frequency value once the microgrid is islanded. 

In future work, we will implement insider attacks as part of our simulation tool. This attacker takes over a trusted and authorized node in the network and intercepts traffic going both directions between the substation and the control center. The attacker can inform the control center that the substation is working properly while conducting a denial of service attack, for example. This attack effectively renders the attacker's action invisible to the control center. By the time the attack is detected, the attacker could have stolen private information or destabilized part(s) of the system.

In addition, we will expand the capability of topology configuration files. We will also add the ability to enable neural network control routing decision to optimize different performance values of the network. We will add the ability to set the protocol used for communication as well as the type of connections, such as point to point, CSMA connections, and/or LTE/\dnp protocols. Finally we will make our tool publicly available as a docker container.

\section*{Acknowledgments}
We would like to thank Md Touhiduzzaman and Burhan Hyder for their valuable feedback on the paper.

\bibliographystyle{ieeetr}
\bibliography{main}

\end{document}